\documentclass[conference]{IEEEtran}
\IEEEoverridecommandlockouts
\usepackage{cite}
\usepackage{soul}
\usepackage{amsmath,amssymb,amsfonts}
\usepackage{algorithmic}
\usepackage{graphicx}
\usepackage{textcomp}
\usepackage{xcolor}
\def\BibTeX{{\rm B\kern-.05em{\sc i\kern-.025em b}\kern-.08em
    T\kern-.1667em\lower.7ex\hbox{E}\kern-.125emX}}
\begin{document}

\title{An Elastic IoT Device Management Platform}

\author{\IEEEauthorblockN{Rakesh Dhakshina Murthy}
\IEEEauthorblockA{\textit{School of Electronic Engineering } \\
\textit{Dublin City University}\\
Dublin, Ireland \\
rakesh.dhakshinamurthy2@mail.dcu.ie}

\and
\IEEEauthorblockN{Mingming Liu}
\IEEEauthorblockA{\textit{School of Electronic Engineering} \\
\textit{Dublin City University}\\
Dublin, Ireland \\
mingming.liu@dcu.ie}
}

\maketitle

\begin{abstract}
With the recent advancement of technologies over the past year, IoT has become a paradigm in which devices communicate with each other and the cloud to achieve various applications in multidisciplinary fields. However, developing, deploying, and experimenting with IoT applications are still tedious, expensive, and time-consuming due to the factors like heterogeneity of hardware and software. This is where an IoT testbed plays a vital role in aiding developers to test their applications without being deploying it to the target environment. In this paper, we present a testbed that is scalable for heterogeneous devices and mainly focused on a small scale and medium scale IoT application. This testbed would be best suited for testing applications which demand robust nature, remote monitoring and control, incorporation of heterogeneous devices, location tracking of devices, and easy troubleshooting with security and internet connectivity concerns. This testbed is also embraced with the feature to work limit access to the internet. A detailed explanation of the design and architecture of the proposed testbed is provided. We also present a conceptual prototype of the testbed and the results obtained on experimenting under various conditions.
\end{abstract}


\section{Introduction}
With recent advancements in the fourth industrial revolution, individuals and organizations, embedded devices have shown a renewed interest towards the Internet of Things (IoT) where millions of heterogeneous devices or things are interconnected. These devices collect real-world data and exchange them over the network \cite{b1}. IoT aims to create a smart environment by making use of the smart things/devices which has sensory and actuating capabilities and transmit the fetched data via the internet. These data could be used in several domains such as energy management, vehicles, transportation, building automation, health care, logistics, power plants, among others \cite{b4a, b4b, b4c, b4d}. It appears that things which are connected to the IoT network are typically heterogeneous and can generate enormous unprecedented and unorganized data flows, especially for time-sensitive applications. The difficulty in handling the data could be addressed by implementing suitable middle-ware between things and cloud serving as a platform for data processing and communication among things as well to the adjacent IoT layers facilitated with interfaces, operating systems, and suitable architecture. Thus, fog nodes come into the picture to complement the toils in IoT \cite{b3}.
Fog nodes are also present at the edge of the network but establishing a group of network resources capable of performing the computation, storage, and networking services between the cloud layer and edge devices/ things. As the things are generally heterogeneous, the data produced by them are also diversified in those scenarios where interoperability and transcoding can be tedious. In this regard, fog nodes perform a key role in time-sensitive applications. 

In general, most IoT applications need to deploy both sensors and actuators in the fields. Before their deployment, these applications usually undergo an appropriate evaluation and testing by using various testing tools. For example, to develop a prototype of a large application involving hundreds of sensors and actuators may not be practical due to economic and operational constraints, especially when a novel IoT protocol is still under consideration and yet to be developed. Moreover setting up the hardware and IoT network elements in the real world is much challenging and often requires expertise knowledge in a particular domain. In order to minimize the expenditure and time for setting up the application, developers usually prefer testbed to test the IoT applications before actually implementing it. In this regard, an IoT testbed serves as a platform for carrying out numerous relentless, transparent, and repetitive testing of IoT applications in different environments \cite{b2, b5}. IoT testbeds are usually designed for the following proposes:
1) To verify the development of the prototype;
2) To verify the operational stability;
3) To perform the test on communication protocols and load;
4) To verify the speed of communication and processing;
and 5) To verify field demonstration (practical test run)\cite{b6}.

Many testbeds have been developed previously targeting different IoT applications (small scale, medium scale, and large scale). In this paper, we discuss the testbed we developed, capable of integrating a large number of heterogeneous things/devices and suitable for small and medium scale applications supporting various protocols. The proposed testbed has a unique feature of working under an \textit{internet-less} environment and considers to have good robustness. In addition, our testbed also comes with a user-friendly GUI support. Both features are seen as the contributions to the paper. The rest of the paper is organized as follows. related works in the literature are described in
Section II. The overall explanation of the testbed are discussed in Section III along with the design procedures and implementation of the prototype in the subsections. Section IV presents the results of the experimentations conducted on the testbed under different circumstances. Finally, the paper is concluded in Section V with future scopes.

\section{Related Work} 
With the advancement in the latest IoT technologies, there is a demand for solutions that require real experimentation for their validation. Several experimental testbeds have been developed over the past years providing multiple services and testing solutions. 

\subsection{FIT loT-LAB}

This test platform has about 2728 low-power wireless nodes and 117 mobile robots. They mainly experiment on
wireless IoT technologies over a large scale from primitive protocols to advanced internet services. They
also offer multi-user scientific tools that are open source. They are deployed across 6 sites each site having
different hardware and software capabilities. These sites are interconnected by a common web portal, CLIs, and rest APIs. As a result, a heterogeneous testing environment is created. ``FIT IoT-LAB” makes use of
open-source to manage their backend operations \cite{b7}.

\subsection{SmartSantander}
The SmartSantander testbed is an advanced testbed which has a large platform and can be used to test entire smart city services and applications. It is targeted for more than 20000 network nodes and these nodes are termed as ``service nodes”. These nodes produce data that can be configured only by the testbed administrators. The end-user does not have access to reprogramming the nodes. However, some nodes with fewer battery constraints are allowed to be programmed by the user allowing the user to test their protocols. IoT and smart city applications are the main targets of the testbed \cite{b8}.

\subsection{Smart CCR IoT}
It is a conceptual framework still under development at the Pheonix metropolitan area aimed at Smart Campus, Cities, and Regions (Smart CCR). This testbed aims to carry out the functioning of the motorization of blue light safety pole on ASU campuses. Each pole module is equipped with a smart data collection unit that can interface with other sensors. The main target of the testbed is to establish a large scale data collection and processing point for future IoT applications \cite{b9}.       

\subsection{AssIUT IOT}
The AssIUT IOT is a remotely accessible testbed and it is based on a five-layer architecture, namely the things layer, sensor layer, nodes layer, communication layer, and the cloud layer. This testbed is widely heterogeneous and capable of incorporating billions of edge devices. This testbed is very complex and requires knowledge in multiple domains to test IoT applications. AssIUT IoT testbed is mainly designed for students and young researchers and it aims to provide sufficient education tools to help them better understand the IoT concepts to build novel IoT application and projects. The testbed mainly makes use of hardware components that can be easily accessible in local markets, and thus it is handy for local researchers to implement the testbed \cite{b10}.

\subsection{SmartICS}
This testbed model mainly focuses on smart buildings and campuses. The testbed primarily makes use of hundreds of indoor environmental sensors, which can be used to actively capture a number of sensor reading quantities related to  air quality, ambient environment, energy consumption and desk occupancy. This testbed has been deployed in the institute of the communication system at the University of Surrey. The researchers make use of two open plane floor on which their devices are fixed on an office floor. The sensor data directly goes to the testbed and is monitored over the application. They are best suited for limited coverage of the area \cite{b11}.

\section{The Proposed IoT Management Platform}
Our proposed testbed is capable of incorporating hundreds of fog nodes and thousands of heterogeneous edge devices and manage them jointly under a common platform. Edge devices are usually sensors or actuators which can be easily controlled by microcontrollers. Smart devices such as Raspberry Pi, Beaglebone, capable of running OS are chosen as fog nodes. The fog node application is deployed on the chosen smart device. Fog nodes communicate with the edge node by embedded protocols like GPIO, Bluetooth, Wifi, UART/USART, I2C. All required computations are carried out in the fog node. The fog nodes are connected to the cloud via a gateway which is, in turn, connects to the internet. Appropriate cloud application is developed to store the data from the edge devices and instruction for the edge devices in their respective cloud database. Testbed management program, which is the key application, is software that could be easily installed on the PC or Laptop. This software is user friendly and comes with catchy GUIs to facilitate simplified user interaction. Once the physical IoT set up is done all the active network elements could be controlled, monitored, tracked, and logically connect and disconnect from the IoT application via testbed application. This testbed is also capable of working in an offline mode i.e. with no access to the internet  \cite{b12}.

\begin{figure} \label{fig1}
	\centering
	\includegraphics[width=3.8in]{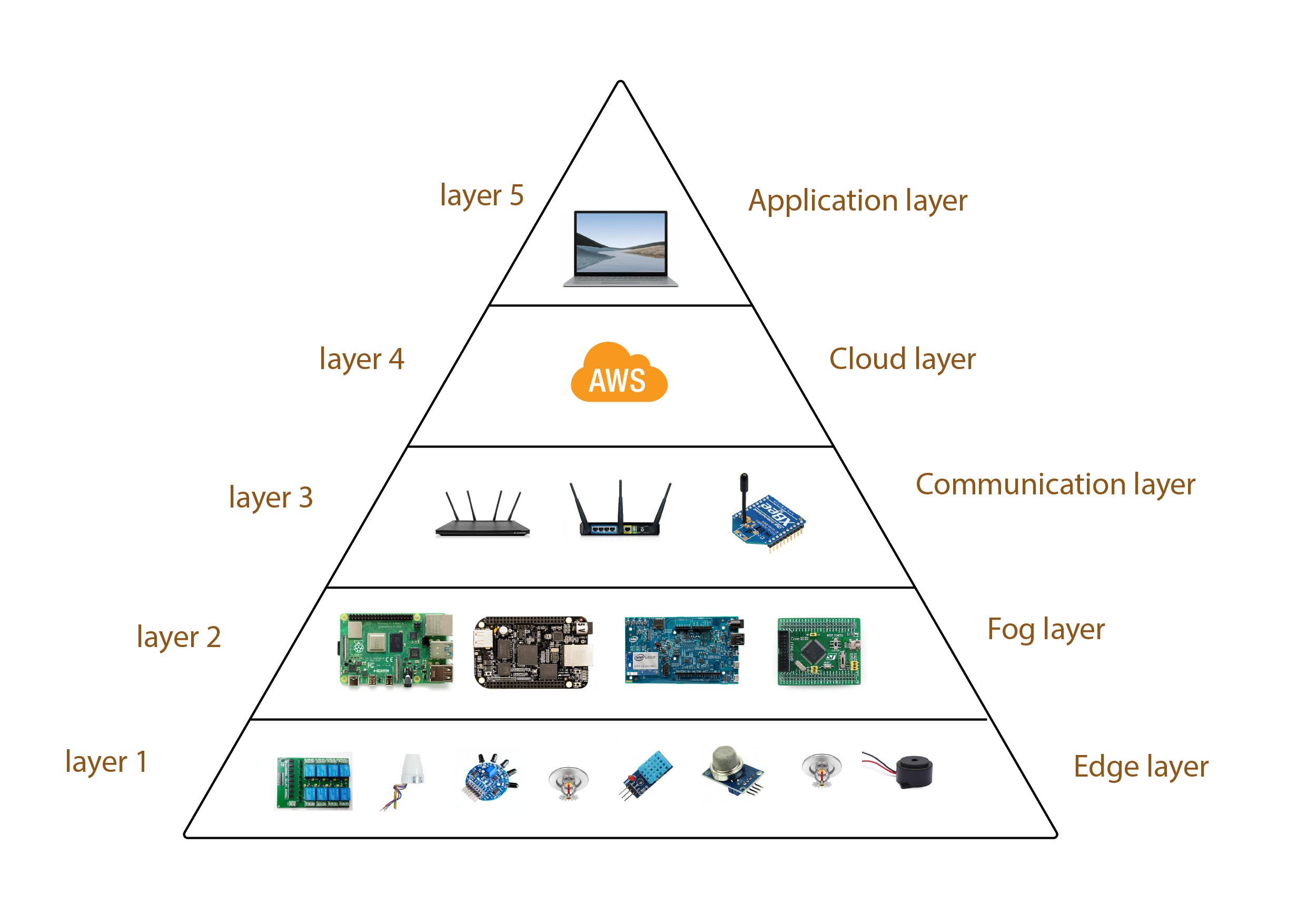}
	\caption{Illustrating multiple layers of a complete IoT system.}
\end{figure}

\subsection{Design and Architecture}
Our IoT testbed is designed to have five stacked layers arranged as shown in the Fig. 1. Each layer plays a specific role in the functioning of the testbed.

\textbf{Layer 1: Edge Layer }
This layer consists of the edge devices namely sensors and actuators which measures and controls the physical parameter of the environment. The desired edge devices are selected according to the requirement of the IoT application and are integrated with the fog nodes. These edge devices are generally heterogeneous and the generated heterogeneous data are sent to the second layer.

\textbf{Layer 2: Fog Layer }
This is the second layer of the testbed. This layer includes the number of distributed fog nodes. To increase efficiency of an IoT system, data from edge layers are processed locally and only the processed data are transmitted to the cloud infrastructure or to other IoT node elements. This helps in reducing the consumption of the bandwidth as well as the response latency.  Fog nodes are designed as smart devices capable of performing computation, communication, and storage. Fog nodes can communicate with the edge devices  using different embedded protocols including such as Bluetooth, I2C, SPI and WiFi. Fog nodes are programmed to carry out three important functions, namely processing the data from their respective edge nodes,  sending the computed data to the Cloud layer via the communication layer, and receiving instructions over the communication layer and function accordingly. 

\textbf{Layer 3: Communication Layer }
This layer acts as a transport layer. The router is the best-opted device for this role. This layer connects the fog nodes to the internet and forwards the appropriate data to the cloud by following relevant protocols and internet standards. Communication between the gateway and the edge devices is usually low power enabled and low rate communication whereas communication with the cloud demands high-speed connectivity. For this purpose, a gateway is also embedded with different protocols such as HTTP/HTTPs, CoAp, MQTT, WebSockets, AMQP, and XMPP \cite{b13}.

\textbf{Layer 4: Cloud Layer }
The Cloud layer can be accessed remotely and it can be used to maximize the resource usage. One of the key functions of the cloud layer is for large data storage. Cloud application receives all the data from the fog nodes and stores them in an organized fashion (e.g. Table). These data are further retrieved from the cloud upon request when desired for further operations.
A cloud application also receives instructions from the application layer and maintains a separate database for the instructions and the details about the fog nodes and their respective edge nodes.

\textbf{Layer 5: Application Layer }
The application layer sits on the top of other layers and it is responsible for the human interaction facility. It is designed as a software that could be installed over any PC/Desktop or Laptop. It serves as a platform where the IoT application could be tested. All the sensors and actuators connected to the IoT application network could be controlled and monitored via this layer remotely. Our testbed application retrieves and sends data and instructions to the cloud database. Other testbed features could be used in this layer. For the testbed to work under the offline mode, it is advised to connect the host machine running the testbed application and the fog nodes to a common gateway.

With these layers in place, the architecture of the proposed testbed is shown in Fig. 2. The testbed operation flow starts with the edge node where the data are sent to their respective fog node for processing. The processed data are sent to the cloud via a gateway and simultaneously instruction stored in the cloud for the fog nodes is retrieved. The testbed application constantly pulls data from the cloud and also updates instructions to the cloud. Under offline mode or sudden failure of the internet connection, the testbed application directly connects to the fog nodes and fetches data from them, and sends the instructions, provided that the host machine and the fog nodes are connected to the common gateway.

\begin{figure*}[h] \label{fig2}
	\centering
	\includegraphics[width=5.0in, height=3in]{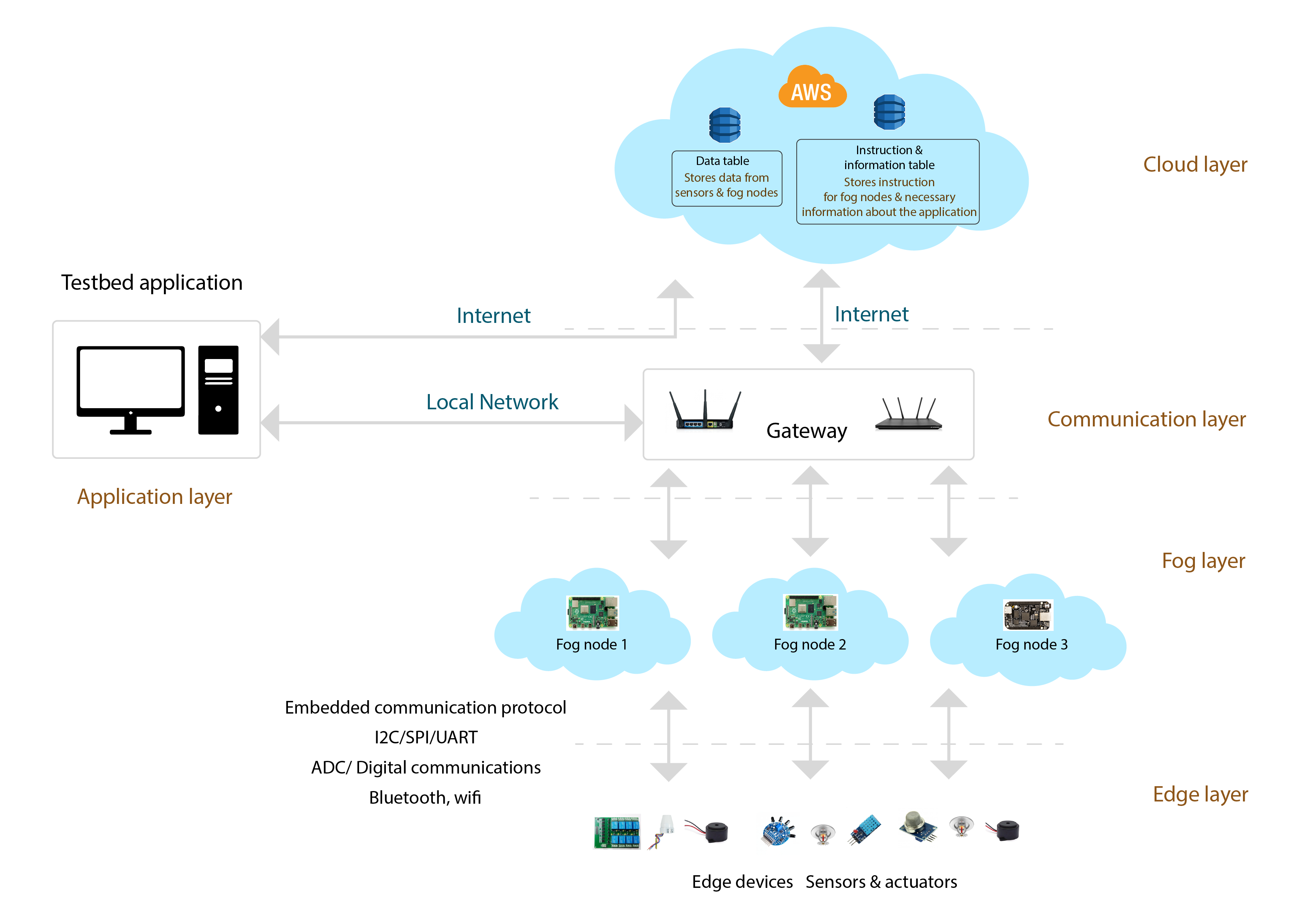}
 	\caption{Architecture and Components of elastic IoT device management platform.}
\end{figure*}

\subsection{Implementation of the Prototype}
In this section, the implementation of the testbed prototype is discussed in the order of the layers. This prototype is designed to have one fog node and four edge node. The physical setup of the prototype is shown in the Fig. 3. We shall first start with the hardware development of the prototype and then discuss the associated software development part.

\begin{figure}[h] \label{fig3}
	\centering
	\includegraphics[width=3.5in, height=3in]{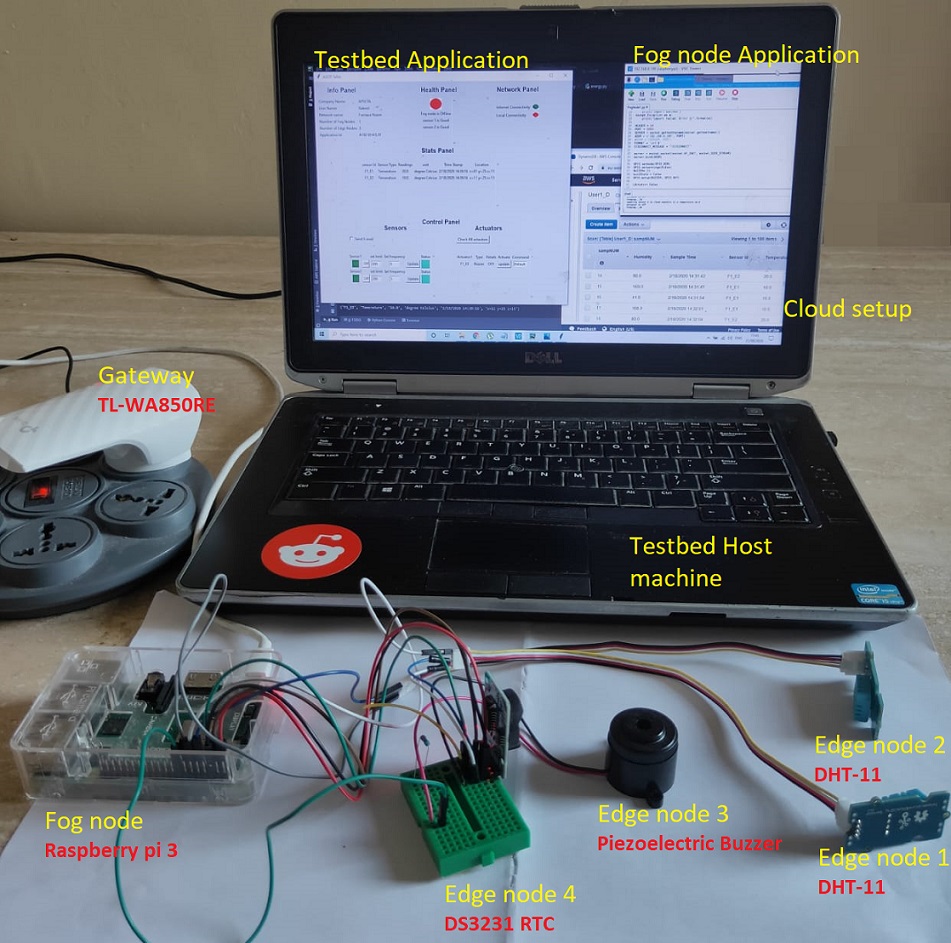}
	\caption{Physical setup of the prototype.}
\end{figure}

\noindent \textbf{Hardware Development:} 

\textbf{Layer 1 :}
Four edge devices chosen for the prototype are two DHT-11 sensors, one DS3231 RTC, and a piezoelectric buzzer. DHT-11 is a well-known, low cost and low power enabled temperature and humidity sensor. It is a capacitive humidity sensor which acts as a thermistor and measures the surrounding air's temperature and humidity. It can be energized by supplying a voltage of 3.3V. For data transmission it makes use of single wire digital protocol transmitting digital data. This sensor was selected because of its worldwide availability, robustness, and easy to handle. DS3231 is a CMOS battery-powered Real-Time Clock (RTC). The DS3231 has been chosen for its high accuracy. It is a RTC that keeps time to ±63 seconds per year (i.e., ±2ppm at 0°C–50°C), and it is widely available in module form at very low cost. It transmits data using I2C or SPI protocol to the controller. The role of RTC in this testbed is to keep track of the timestamp at which the sensor data are uploaded as well as to maintain a local clock.  A piezoelectric buzzer is a simple electronic device that produces sound or tone when powered and acts as an actuator (alarm). The intensity of the sound increases with the increase with the power supplied to the buzzer. The power rating ranges from 3.3v to 9V.  

\textbf{Layer 2 :}
For the development of the prototype Raspberry Pi 3 is selected as a fog node. It is a credit card-sized computer that has 1.2 GHz ARM Cortex-A53 CPU supporting 64bits architecture with a 32GB SD card enabled storage feature. It has four USB ports, one HDMI port, and one ethernet port. Basic accessories like monitor, keyboard, and mouse could be connected. It supports Debian and has interface options like GPIOs, SPI, UART, I2C, and built-in Wifi support. The edge nodes are connected to the fog node using the interface options. The Wifi support is used to connect the fog node to the gateway which access to the public internet.

\textbf{Layer 3 :}
In this prototype, TP-LINK's TL-WA850RE, a Wifi extender connected to a router with an internet connection is used as a gateway. It has an Ethernet port that allows the TL-WA850RE to act as a wireless adapter to turn a wired device into a wireless one. 

\noindent \textbf{Software Development:} 

\textbf{Layer 2 :} Raspberry Pi is used as a fog node with Raspbian OS. AWS SDK is installed on the fog node device. Firmware for fog node is developed by using python programming languages and AWS boto3 services. This application is later deployed on the fog node. 

\textbf{Layer 4 :} For the cloud set-up, AWS is used. In particular, the AWS DynamoDB  service is used to create a database table for storing the data and instruction of the edge devices and fog nodes. Amazon DynamoDB provides predictable and high speed performance with seamless scalability. It is a fully managed NoSQL database service. 

\textbf{Layer 5 :} The testbed application is a software developed to aid the users for efficient handling of the testbed. This testbed software is developed using python programming language and later converted into an executable file. Some important frameworks used in the development are boto3 and Tkinter. 
Boto3 is an SDK developed by AWS for easy implementation and management of AWS services using python. Tkinter is a python building tool for GUI. Tkinter calls are translated into Tcl commands which are fed to this embedded interpreter, thus making it possible to mix Python programming language and Tcl in a single application.

\section{Experimentation and Observation}

This section discusses the behaviour of the testbed under different experimental scenarios demonstrating the key features of the proposed testbed compared to others. The below figure shows the interface of the testbed software.

\begin{figure}[h]
	\centering
	\includegraphics[width=3.3in, height=3in]{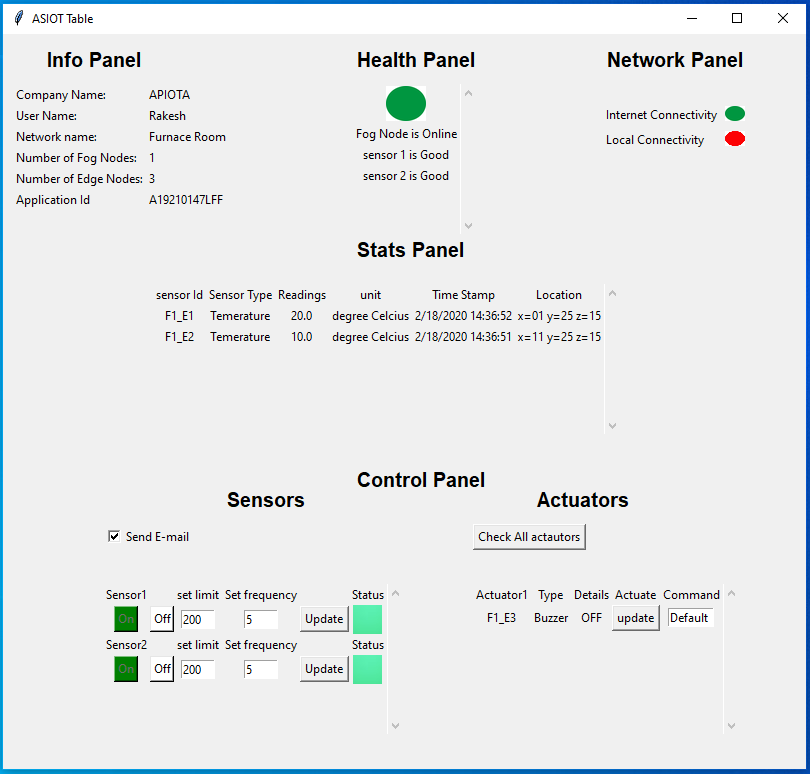}
	\caption{User interface of the elastic IoT device management platform.}
\end{figure}

\subsection{Overall description of the testbed} Different user interfaces of the proposed testbed application are shown in Fig. 4. 
\begin{itemize}
	\item General information about the user and the application being tested and the field of test area are displayed under the info panel.
	\item The health of all the network elements and fog nodes of the application is displayed in the health panel. This feature is added to ensure the proper working of the testbed elements. 
	\item The proposed testbed is capable of working with or without the internet. The current network in which the testbed works is indicated in this panel.
	\item The live statistical values of all the sensors used in the application are displayed on the stats panel. Details such as sensor id, location of the sensor, sensor reading, and timestamp could be referred from this panel.
	\item Edge devices (sensors and actuators) can be controlled by the options provided in the control panel.
\end{itemize}

\subsection{Testbed Normal Workflow}
\begin{itemize}
	\item Under the normal working condition, the live data from the sensors could be monitored on the stats panel along with the location and timestamp.
	\item The sensors could be identified by using their respective sensor id.
	\item Under the control panel, sensors could be logically connected or disconnected from the network by clicking on the 'On/Off' button.
	\item The user is also provided with an option to set an ideal working range for the sensor. When the sensor reading exceeds the pre-defined limit, the respective indicator of the sensor turns red from green indicating abnormality. 
	\item Provision is provided to modify the frequency at which the sensors pushes data to the cloud.
	\item The user is also provided with an additional feature of sending E-mail to the operator when the sensor exceeds the limit. This could be enabled by checking the `send E-mail' checkbox.
	\item When the sensor exceeds the limit appropriate actuators are automatically turned on. Under the actuator section, features are provided to turn on the actuator manually as well as to send commands to the actuators. Since the actuator is a buzzer in our prototype set-up, a command such as duration or tone of the buzzer can be set.
	\item 'Check All actuators' button enables the user to check simultaneously all the actuators in the field manually.
\end{itemize}

\subsection{Failure of Edge/Fog Nodes}
\begin{itemize}
	\item  There are many reasons an edge/fog node could fail especially due to failure in power. Any failure of sensors or actuators in the IoT network is reflected immediately on the health panel.
	\item The regular functioning of the testbed is not affected by the failure of devices. The faulty node or device is indicated precisely along with the location on the testbed application.
	\item To experiment this scenario, a faulty edge node is chosen as sensor 1. The behaviour of the testbed is observed. The system continues to function normally and the response of the testbed is shown in the Fig. 5.
\end{itemize}

\begin{figure}[h]
	\centering
	\includegraphics[width=3.6in]{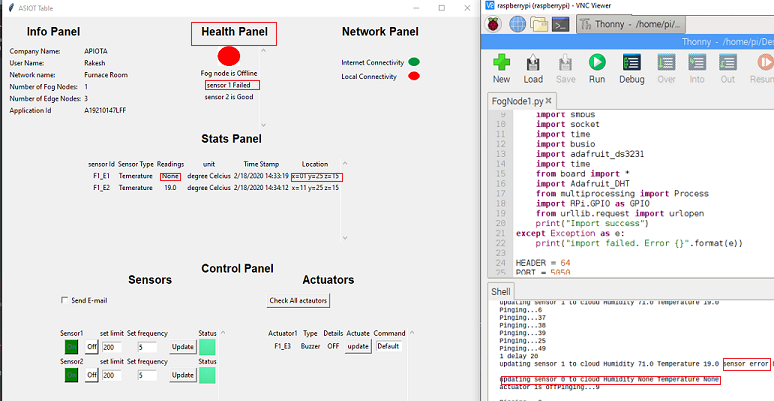}
	\caption{Testbed working with failed edge device}
\end{figure}

\subsection{Testbed in Offline Mode}
\begin{itemize}
	\item This unique feature could be useful in a situation where proper internet could not be provided to the fog node or the host machine running the testbed application.
	\item Fog node firmware is developed in such a way each fog node acts as an individual server and the testbed application behaves as a client. By using a socket communication protocol, the fog node and the machine running the testbed application are connected to a common gateway through which the application sends and receives data to and from the fog node in the form of a data stream.
	\item When this feature is enabled, it is indicated in the network panel.  
	\item In case of a sudden failure of the internet connectivity of the fog node or testbed application, the testbed immediately switches to offline mode and continues its working, provided they are connected to the common gateway. This feature increases the robustness of the testbed. If not, then the appropriate failure reason is indicated on the health panel. 
	\item To further experiment this feature, the internet connectivity to the gateway is disconnected. As can be seen in Fig. 6, the testbed immediately shifts to the offline mode and the testbed application continues to interact with the fog node flawlessly. 
\end{itemize}

\begin{figure}[h]
	\centering
	\includegraphics[width=3.6in]{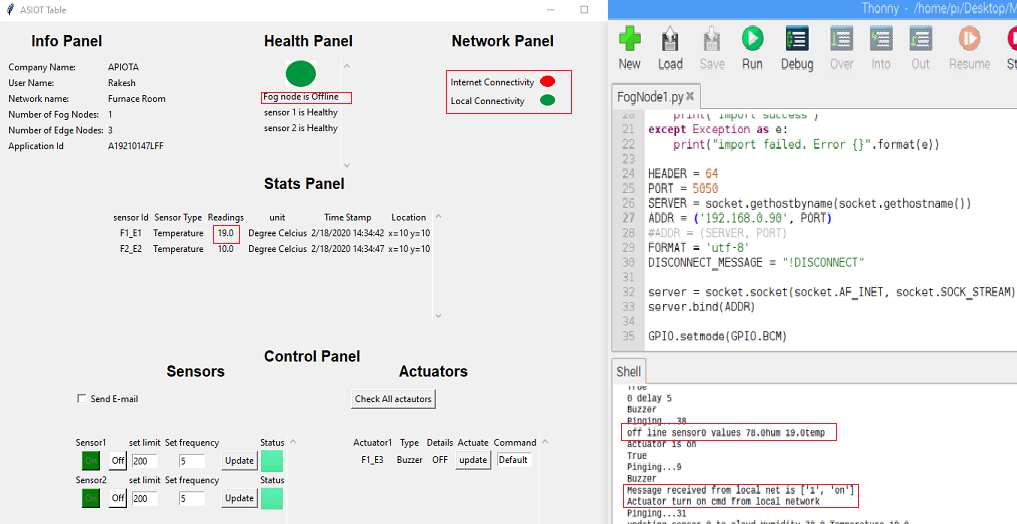}
	\caption{Testbed working in offline mode.}
\end{figure}

\subsection{Unauthorized User Access}
\begin{itemize}
	\item As aforementioned, the testbed can work both with only local networks and with the internet, thus it is possible for unauthorized users to breach into the system. 
	\item As a preventive measure, the fog node's firmware is programmed in a way that any external client tries to access the fog node number of active clients changes. Any anonymous user trying to break into the system can then be detected immediately. 
	\item Additionally, all data transfer using the testbed is encrypted and protected with an encryption key.
	\item Cloud level security features are provided by AWS services. DynamoDB encrypts at rest all user data stored in tables, indexes, streams, and backups using encryption keys stored in AWS Key Management Service (AWS KMS). This provides an additional layer of data protection by securing users' data from unauthorized access to the underlying storage.
\end{itemize}

\section{Conclusion}
In this paper, we presented a general overview of our proposed testbed, which serves as an elastic platform for heterogeneous devices mainly focused on experimenting and developing IoT applications. Hardware and software architecture have also been discussed for the prototype developed. This testbed could be used for small scale and medium scale applications mainly for developing IoT protocols. This testbed does not require the internet every time for its functionality and its very robust. This testbed could also be shared by multiple users simultaneously in a secure way and capable of withstanding high load. Tangible results obtained from running the prototype of the testbed are also shown and discussed in this paper. The proposed testbed shows high robust behaviour and very low possibilities of failure even under the internet failure. It has a simple yet an efficient architecture and components making it very reliable and realisable with a better interface when compared to other developed testbeds in the literature. As part of our future work, this testbed could be further improvised to adopt a machine learning model which can provide predictive analysis of the sensors in real time.

\section*{Acknowledgement}

The authors acknowledge the support from the School of Electronic Engineering at Dublin City University to conduct this project. 

\bibliographystyle{ieeetran}
\bibliography{References}

\end{document}